\documentclass[twocolumn,showpacs,aps,prb,preprintnumbers,amsmath]{revtex4}
\usepackage{amssymb,amsmath}
\usepackage{graphicx}
\usepackage{bm}
\begin{document} 

\title{Vogel-Fulcher freezing in relaxor ferroelectrics}

\author{R. Pirc and R. Blinc}

\affiliation{Jo\v zef Stefan Institute,  P.O. Box 3000, 1001 Ljubljana, 
Slovenia}

\date{\today}

\begin{abstract}
A physical mechanism for the freezing of 
polar nanoregions (PNRs) in relaxor ferroelectrics is presented.
Assuming that the activation energy for the reorientation of a 
cluster of PNRs scales with the mean volume of the cluster, 
the characteristic relaxation time $\tau$ is found to diverge as the 
cluster volume reaches the percolation limit. Applying the mean field 
theory of continuum percolation, the familiar Vogel-Fulcher equation
for the temperature dependence of $\tau$ is derived.
\end{abstract}\pacs{77.22.Gm, 77.84.Dy, 67.40.Fd}
\maketitle
~~~~

Relaxor ferroelectrics (relaxors) have long been attracting considerable
attention in view of their unique physical properties.\cite{S1} A key 
feature of relaxors is the appearance of a broad temperature and  
frequency dependent maximum of the dielectric permittivity 
$\epsilon(T,\omega)$ as well as the absence of long range ferroelectric 
order in zero field at any temperature.\cite{C1} The high value 
of the quasistatic dielectric constant $\epsilon$ in a broad 
temperature range around the peak temperature $T_m$, and especially
the giant piezoelectric effect\cite{KBP,RC} make 
relaxors attractive for various 
technological applications. Another characteristic relaxor property 
is the extremely slow relaxation below $T_m$, which 
signals the onset of relaxor freezing.\cite{VJCW} The characteristic 
relaxation time diverges at the freezing temperature $T_0$ according 
to the well known Vogel-Fulcher (VF) relation\cite{V1,F1}
\begin{equation} 
\tau = \tau_0 \exp[U/k(T - T_0)],
\label{VF}
\end{equation}
where $\tau_0$ represents the inverse attempt frequency, 
$U$ the activation energy, and $T_0$ the VF or freezing temperature. 
For obvious reasons, Eq.\ (\ref{VF}) only makes sense for $T > T_0$. 

The above empirical VF law (\ref{VF}) has been experimentally observed  
in a variety of other systems such as supercooled organic liquids, 
spin glasses, polymers etc. Although many theoretical ideas about the
origin of the VF law have been proposed in the 
past,\cite{C2,C3,W1,SW,LSB,S2,T1} a derivation of Eq.\ (\ref{VF})
at the mesoscopic level is still lacking.

Much of the experimental and theoretical research on relaxors
has been focused on compositionally disordered perovskites such as
PbMg$_{1/3}$Nb$_{2/3}$O$_3$ (PMN) and related compounds.\cite{S1} 
Since only disordered ferroelectric systems show relaxor behavior, 
it is clear that disorder and/or frustration are key factors 
in producing a relaxor state. In particular, charge fluctuations 
are responsible for the formation of polar nanoregions (PNRs) below the 
so-called Burns temperature $T_d$.\cite{BD} In PMN, 
for example, one has $T_d \sim 600$ K,
while the dielectric maximum occurs at $T_m \sim 260$ K in the 
quasistatic limit. The PNRs can be regarded as a network of randomly
interacting dipolar entities with the corresponding statistical distributions 
of their size and dipolar strength. The resulting collective state is 
reminiscent of a magnetic spin glass---or rather an electric dipolar 
glass---and it has been shown that its static properties can be described 
in terms of the spherical random bond--random field (SRBRF) 
model of relaxor ferroelectrics.\cite{P1,P2} 
Experiments performed by neutron scattering\cite{VKN,JDL}
and NMR techniques\cite{BLZ} indicate
that the average size of PNRs increases with decreasing temperature
and saturates below $T_m$, suggesting the possibility of a 
percolation-type transition into a frozen relaxor state.\cite{S1,B1}

Relaxor dynamics is characterized by a broad distribution of 
relaxation times $g(\log\tau)$, and it appears that the
freezing process is associated with the divergence of the longest
relaxation time in $g(\log\tau)$.\cite{VJCW,LKFP} A typical empirical
method used to analyze the dielectric permittivity $\epsilon(\omega,T)$
is to consider the dielectric maximum temperature $T = T_m$ as a function
of frequency $\omega$. One may then define a relaxation time 
$\tau =1/\omega$, which is found to satisfy the above 
Vogel-Fulcher (VF) relation (\ref{VF}). For example, in PMN oriented 
along $[001]$, the experimental parameter values determined in this manner 
are: $\tau_0 = 10^{-12}$~s, $U/k = 911$~K, and $T_0 = 217$~K.\cite{VCW}
On the other hand, the longest relaxation time $\tau_{max}$ was found
to obey the VF relation (\ref{VF}) with similar parameter values, i.e., 
$\tau_0 = 4.3 \times 10^{-11}$~s, $U/k = 970$~K, 
and $T_0 = 224$~K.\cite{LKFP}

The purpose of the
present work is to present a simple physical mechanism for the 
VF-type relaxation process in relaxor ferroelectrics. We adopt
a qualitative physical picture of the relaxor state below $T_d$ 
based on a network of PNRs embedded in a highly polarizable medium.\cite{S1}
One can imagine that the medium consists of a number of fluctuating  
reorientable dipoles and/or small size polar clusters. The first
possibility corresponds to dipolar glasses and order-disorder-type
relaxor ferroelectrics, whereas the second one applies to displacive-type 
relaxors; however, intermediate cases may as well exist.
Each PNR will polarize the medium within a space region 
bounded by the correlation radius $r_c$.
As the temperature is lowered,  $r_c$ is expected to increase;
this process will continue until a number of PNRs of similar size 
start to merge, thus forming a polarization cluster. Eventually, 
freezing will occur due to the growth of both the size of PNRs and 
the correlations between them.\cite{VCW}

Let us now consider a PNR with a core radius 
$r_0$ and assume that the polarization cloud associated 
with it can be described by a power-law radial dependence, 
\begin{equation} 
\vec{P}(r) = \vec{P}_0 (r_0/r)^3; \;\; r > r_0,
\label{Pi}
\end{equation}
and $\vec{P}(r) = \vec{P}_0$ for $r\le r_0$. 
The local electric field at some distance $r$ is proportional to 
the Lorentz field\cite{VG}
\begin{equation} 
\vec{\mathcal{E}} = \frac{\varphi}{3\epsilon_0}\vec{P}(r),
\label{Er}
\end{equation}
with a local field correction factor $\varphi = O(1)$. 
The field $\vec{\mathcal{E}}$ couples to dipolar fluctuations 
of the surrounding medium and at a distance $r$ induces an electric 
dipole moment  
\begin{equation} 
\vec{p} = \alpha \vec{\mathcal{E}},
\label{pi}
\end{equation}
which is proportional to the polarizability $\alpha$. 
For order-disorder relaxors one has $\alpha \simeq \mu^2/kT$,
where $\mu$ is the strength of the fluctuating dipole moment. 
Similarly, in the displacive case we can write 
$\alpha \simeq e^{*2}/(M\omega_0^2)$, where $e^*$ is an effective 
charge, $M$ the reduced mass, and $\omega_0^2 \simeq a_0 kT$ the 
frequency of a renormalized quasi harmonic mode, which becomes 
unstable at zero temperature. The associated change of the 
electrostatic energy is given by
\begin{equation} 
\delta E = - \frac{1}{2} \alpha {\mathcal E}^2.
\label{dE}
\end{equation}

If $|\delta E| > kT$, the thermal fluctuations will be too weak 
to destroy the correlations between the dipole and the PNR and 
a bound state will exist. 
The correlation radius $r_c$ then corresponds to the limiting 
distance for which $|\delta E| \simeq kT$. Combining Eqs.\ (\ref{dE}), 
(\ref{Pi}), and (\ref{Er}) we obtain 
\begin{equation} 
r_c^3 = r_0^3 \frac{T_d^*}{T},
\label{rc3}
\end{equation}
where $T_d^* = \varphi \mu P_0/3\sqrt{2}k\epsilon_0$ for
order-disorder relaxors, and similarly 
$T_d^* = \varphi e^*P_0/3\sqrt{2Ma_0}k\epsilon_0$
for the displacive case. Formally, we can require that $r_c \to r_0$ as
$T \to T_d$, implying that $T_d^*$ is of the order $\sim T_d$.

It follows that the correlation radius of each PNR in each case 
scales with temperature $t \equiv T/T_d^*$ as $r_c \sim t^{-1/3}$. 
Similarly, the correlation volume $v_c = 4\pi r_c^3/3$ scales as 
$\sim 1/t$, and the PNR dipole moment 
$p_c = (4\pi/3)\int_0^{r_c} P(r) r^2 dr$ as 
$p_c -  p_0 \sim \vert \log t \vert$, where 
$p_0 = P_0 4\pi r_0^3/3$ is the core dipole moment.

As the temperature is lowered, the PNRs will grow in size and gradually 
start forming a connected polarization cluster. If $n$ is the 
concentration of PNRs, they occupy a volume fraction 
$\eta = 4\pi n r_c^3/3 = 4\pi n T_d^*/3T$. When $\eta$ reaches a 
threshold value $\eta_p$, the PNRs will merge into an infinite 
cluster---a familiar concept from the theory of percolation.\cite{I1} 
The temperature at which the infinite cluster appears is therefore 
$T_p = 4\pi n T_d^*/3\eta_p$.
The percolation threshold for hard spheres on a lattice in three 
dimensions ($d=3$) is $\eta_p \simeq 0.35$, and $\eta_p \simeq 0.294$ 
for randomly overlapping spheres,\cite{I1} whereas $\eta_p = 1/3$
for random hard spheres in the effective medium approximation.\cite{T2} 
Experiments on PMN\cite{JDL} show that the volume fraction of
PNRs saturates at $\sim 0.3$ below $T \sim 15$~K.

The theory of continuum percolation\cite{D1} predicts that the mean
cluster volume $v$ increases as $\eta$ approaches $\eta_p$ 
according to a power law  $v \sim (\eta_p - \eta)^{-\gamma}$, 
or explicitly
\begin{equation} 
v = v_0 (1 - \eta/\eta_p)^{-\gamma},
\label{V}
\end{equation}
where $v_0$ is the critical amplitude. Assuming that at high
temperatures $T \sim T_d$ the mean cluster volume reduces to the
average core volume, we obtain an estimate for the amplitude,
i.e., $v_0 \sim 4\pi  r_0^3/3$. 

In the mean field case, which is applicable to systems with infinite 
effective dimensionality, one has $\gamma = 1$. This may be a 
reasonable value for the present system, since the PNRs behave as 
a fully connected random-site network in view of the long range 
character of dipolar interactions. Using the above expressions for 
$\eta$ and $\eta_p$ we can rewrite Eq.\ (\ref{V}) as
\begin{equation} 
v = \frac{v_0}{(1-T_p/T)}.
\label{VT}
\end{equation}

As the volume of the cluster grows, it becomes increasingly difficult
for its total dipole moment to change direction, and eventually the
reorientation will be suppressed completely as the mean cluster
size reaches the percolation limit. The relaxation time $\tau$ for
the reorientation of the dielectric polarization is usually described 
by the Arrhenius law $\tau = \tau_0 \exp(U/kT)$, where the
activation energy $U$ is determined by the potential barrier for 
the relaxing particle to jump out of the potential well. 
For magnetic clusters in disordered magnetic materials, N\' eel\cite{N1} 
suggested that the activation energy could be written as
the product of mean cluster volume $v$ and an anisotropy factor $Q_{an}$,
\begin{equation} 
\tau = \tau_0 \exp (vQ_{an}/kT).
\label{NE}
\end{equation}
As already noted by several authors\cite{VJCW,S1,B2} an analogous 
relation should be applicable to PNRs in relaxors.
Inserting the mean cluster size $v$  from Eq.\ (\ref{VT}) 
into Eq.\ (\ref{NE}) we immediately obtain
\begin{equation} 
\tau = \tau_0 \exp [v_0Q_{an}/k(T-T_p)].
\label{Vp}
\end{equation}
This result has precisely the form of the VF equation (\ref{VF}) with
$U = v_0Q_{an}$ and $T_0 = T_p$. 

The anisotropy constant $Q_{an}$ in N\' eel's formula (\ref{NE}) has
the dimensionality of energy density. Its magnitude can be estimated by
assuming that the core radius $r_0$ is of the order, say,
$\sim 2$~nm. Using the value $U/k = 970$~K for PMN\cite{LKFP} we 
thus find $Q_{an} \sim 2.3 \times 10^{-3}$~eV/(nm)$^3$.

The VF relation (\ref{Vp}) for the relaxation of dielectric polarization 
in relaxor ferroelectrics has been derived here on the basis of a 
plausible power-law model for the polarization distribution within
the PNR. The principal mechanism responsible for relaxor freezing 
appears to be the growth and percolation of PNR clusters\cite{S1}
culminating in the formation of an infinite cluster. Alternatively,
the system could be described in terms of random normal modes, which 
are determined by the eigenstates and eigenvalues of the random
interactions between PNRs;\cite{P3} however, the relaxation time
entering the equations of motion for these modes must again be associated
with the growth of the PNRs discussed above and should, therefore, obey 
the same VF relation. Thus, according to our present model, 
the two physical pictures, namely, that of cluster growth and 
of the freezing of local modes,\cite{VJCW,C4}  
are essentially two parts of the same general scenario.

It should be noted that the above results are independent of any 
specific spin glass-type static model such as the SRBRF model. 
Of course, the SRBRF model remains applicable to true 
{\it static} phenomena, for example, the temperature dependence 
of the quasistatic dielectric response as observed in a 
field-cooled (poled) sample.

The divergence of $\tau$ at $T_p$ does not imply that the motion of
all PNRs is completely frozen for $T < T_p$. Namely, the complex 
dielectric permittivity $\epsilon(\omega,T)$ remains finite
below $T_p$, indicating that some degrees of freedom are 
still active at low temperatures. In fact, as already noted above, 
only $\sim 30 \%$ of the PNRs are involved in the formation of  
the infinite cluster. Smaller PNRs in the remaining space may continue 
to undergo a similar process of growth and percolation, suggesting that
at any temperature below $T_p$ an analogous freezing mechanism 
may apply. One can thus introduce a probability distribution 
of VF temperatures $w(T_p)$ in the interval $0 < T < T_p$ which could, 
in principle, be transformed into a nontrivial distribution of
relaxation times. A simple linear shape of $w(T_p)$\cite{P3} 
then reproduces the main physical features, 
i.e., the finite value of $\epsilon(\omega,T)$ at temperatures 
below $T_p$ and a frequency dispersion of its real and imaginary 
parts, although the agreement with the observed behavior 
of $\epsilon(\omega,T)$ is---admittedly---only qualitative.

At present, it is not yet clear whether the above approach 
is applicable in a straightforward manner to analogous systems 
such as random ferromagnets and spin glasses, supercooled organic 
liquids, structural glasses, etc. In magnetic systems, the magnetic dipolar
interactions are much weaker than the short range exchange interactions,
which are responsible for the formation of superparamagnetic
clusters. However, in spin glasses the RKKY interactions 
actually fall off as $\sim r^{-3}$ up to some oscillatory prefactor, 
and magnetic dipolar interactions are believed to be relevant in 
some spin glass systems.\cite{FH} Thus the present model may serve 
as a convenient starting point for these systems. Meanwhile, 
in relaxor ferroelectric polymers, the above physical picture seems
to be applicable without serious limitations.\cite{P2} For  
structural glasses and supercooled organic liquids, the possibility 
of elastic deformations and their interactions playing a leading role 
in the growth of correlated clusters should be explored.

In conclusion, we have derived the Vogel-Fulcher (VF) relation in 
relaxor ferroelectrics by introducing a mesoscopic mechanism for 
the growth of PNRs. The basic idea is that thermodynamic stability 
of the polarization density with a power-law distribution 
$P(r) \sim r^{-3}$ entails a temperature dependence of the correlation 
radius $r_c \sim T^{-1/3}$. Thus a cluster of PNRs is formed on 
lowering the temperature, and its mean volume $v$ increases until 
the percolation limit is reached at some temperature $T_p$. 
Using the mean field theory of continuum percolation we have shown 
that $v \sim (1 - T_p/T)^{-1}$, and since according to N\' eel the 
activation energy $U$ for the reorientation of the cluster polarization 
is proportional to $v$, the VF relation follows immediately.

This work was supported by the Slovenian Research Agency through 
Research Programs P1-0044 and P1-0125. Stimulating discussions with
Z. Kutnjak and V.S. Vikhnin are gratefully acknowledged.



\end{document}